\documentclass{article}
\usepackage{spconf,amsmath,graphicx}
\usepackage{subfigure}
\usepackage{mathrsfs}
\usepackage{float}
\usepackage{hyperref}
\usepackage{url}
\usepackage{multirow}
\usepackage{graphicx}
\usepackage{xcolor}

\usepackage[ruled,vlined]{algorithm2e}
\title{Seen and Unseen emotional style transfer for voice conversion \\ With a new emotional speech dataset}
\name{Kun Zhou $^{1}$,\thanks{\textbf{Codes \& speech samples:} \url{https://kunzhou9646.github.io/controllable-evc/}} Berrak Sisman $^{2}$, Rui Liu $^{2, 1}$, and Haizhou Li $^{1}$}
\address{
$^{1}$ Department of Electrical and Computer Engineering, National University of Singapore\\
$^{2}$ Information Systems Technology and Design, Singapore University of Technology and Design}

\begin{document}
\ninept
\maketitle
\begin{abstract}
Emotional voice conversion aims to transform emotional prosody in speech while preserving the linguistic content and speaker identity. Prior studies show that it is possible to disentangle emotional prosody using an encoder-decoder network conditioned on discrete representation, such as one-hot emotion labels. Such networks learn to remember a fixed set of emotional styles. In this paper, we propose a novel framework based on variational auto-encoding Wasserstein generative adversarial network (VAW-GAN), which makes use of a pre-trained speech emotion recognition (SER) model to transfer emotional style during training and at run-time inference. In this way, the network is able to transfer both seen and unseen emotional style to a new utterance.   We show that the proposed framework achieves remarkable performance by consistently outperforming the baseline framework. This paper also marks the release of an emotional speech dataset (ESD) for voice conversion, which has multiple speakers and languages.


\end{abstract}

\section{Introduction}
Speech conveys information with words and also through its prosody. Speech prosody can affect the syntactic and semantic interpretation of an utterance (\textit{linguistic prosody}), and also displays one's emotional state (\textit{emotional prosody}) \cite{hirschberg2004pragmatics}. Emotional prosody reflects the intent, mood and temperament of the speaker and plays an important role in daily communication \cite{arnold1960emotion}. Emotional voice conversion is a voice conversion (VC) technique, which aims to 
transfer the emotional style of an utterance from one to another. Emotional voice conversion enables various applications such as expressive text-to-speech (TTS) \cite{liu2020teacher} and conversational agents.

Emotional voice conversion and speech voice conversion \cite{sisman2020overview} differs in many ways. Speech voice conversion aims to change the speaker identity, whereas emotional voice conversion focuses on the emotional state transfer. Traditional VC research includes modelling spectral mapping with statistical methods such as Gaussian mixture model (GMM) \cite{toda2007voice}, partial least square regression \cite{helander2010voice} and sparse representation \cite{sisman2019group}. Recent deep learning approaches such as deep neural network (DNN) \cite{chen2014voice}, recurrent neural network (RNN) \cite{nakashika2014high} and generative adversarial network (GAN) \cite{sismanstudy} have advanced the state-of-the-art. We note that these frameworks have motivated the studies in emotional voice conversion. 



Early studies on emotional voice conversion handle both spectrum and prosody conversion with GMM \cite{tao2006prosody} and sparse representation \cite{aihara2014exemplar}. Recent deep learning methods, such as deep belief network (DBN) \cite{luo2016emotional}, deep bi-directional long-short-term memory (DBLSTM) \cite{ming2016deep}, sequence-to-sequence \cite{robinson2019sequence} and rule-based model \cite{xue2018voice} have shown the effectiveness on emotion conversion. We note that these frameworks are relied on the parallel training data. However, such data 
is not widely available in real-life applications.

There have been studies on deep learning approaches for emotional voice conversion that do not require parallel training data, such as cycle-consistent adversarial network (CycleGAN)-based \cite{Zhou2020, shankar2020non} and autoencoder-based frameworks \cite{gao2019nonparallel,zhou2020converting}.
However, they are typically designed for a fixed set of conversion pairs.  In this paper, we would like to propose a novel technique, that is referred to as emotional style transfer (EST), that learns to transfer an emotional style to any input utterance. The emotional style is given to the network as a control condition. Therefore, the network supports one-to-many emotion conversion and for unseen emotion at run-time.

Auto-encoder is a suitable computational model that allows for the control of output generation through the latent variables~\cite{hsu2017voice}. Recent studies~ \cite{zhou2020converting} on disentangling and recomposing the emotional elements in the speech with VAW-GAN  represent one of the successful attempts in emotional voice conversion. 
However, emotional prosody is the result of the interplay of multiple signal attributes, hence it is not easy to define emotional prosody by a simple labeling scheme \cite{luong2017adapting}. In TTS studies, there are techniques to learn a latent prosody embedding, i.e. style token, from the training data, in order to predict or transfer the speech prosody \cite{wang2018style,skerry2018towards}. 
These studies motivate us to investigate the use of deep emotional features that reflect the global speech style and describe the emotional prosody in a continuous space.

In this paper, we propose an emotional style transfer framework based on VAW-GAN. We use deep emotional features as a condition to enable the encoder-decoder training for seen and unseen emotion generation. Furthermore, we release an EVC dataset that consists of multi-speaker and multi-lingual emotional speech. We aim to tackle the lack of open-source emotional speech data in voice conversion research community. This dataset can be easily applied to other speech synthesis tasks, such as cross-lingual voice conversion and emotional TTS.

The main contributions of this paper include:  1) we propose to build a one-to-many emotional style transfer framework that does not require parallel training data; 2) we propose to use SER that is pre-trained with  publicly available large speech corpus to describe the emotional style;
3) we propose to disentangle and recompose the emotional elements through deep emotional features; and 4) we release a multi-lingual and multi-speaker emotional speech corpus, denoted as ESD, that can be used for various speech synthesis tasks. To our best knowledge, this paper is the first reported study on emotional style transfer for unseen emotion.

This paper is organized as follows: In Section 2, we motivate our study and analyze the deep emotional features. In Section 3, we introduce the proposed one-to-many EST framework. In Section 4, we report the experiments. Section 5 concludes the study.

\section{Analysis of Deep Emotional Features}

The computational analysis of emotion has been the focus of SER~\cite{schuller2020review}. Recent advances of deep learning have led to a shift from traditional human-crafted representations of acoustic features, to the deep features automatically learnt by neural networks \cite{schuller2020review,latif2020deep}. Deep features are data-driven, less dependent on human knowledge and more suitable for emotional style transfer~\cite{schuller2020review}.

Emotional prosody is prominently exhibited in emotional speech databases, which can be characterized by discrete categories, such as Ekmans's six basic emotions \cite{ekman1992argument}, and continuous representation, such as Russell's circumplex model \cite{russell1980circumplex}. Recent studies seek to characterize emotions over a continuum rather than a finite set of discrete categorical labels with the feature representation learnt by deep neural networks in a continuous space \cite{kim2019dnn}. Following the findings in speech analysis, studies are conducted to use emotional prosody modelling to improve the expressiveness of speech synthesis systems \cite{gao2020interactive,um2020emotional,liu2020expressive}. In \cite{gao2020interactive}, an emotion recognizer is used to generate the style embedding for speech style transfer. Um et al. \cite{um2020emotional} apply the style embedding to a Tacotron system with the aim to control the intensity of emotional expressiveness. These successful attempts have revealed the fact that deep emotional features serve as the excellent prosody descriptor, which motivates this study.

\begin{figure}[t]
\centering
\subfigure[Data in left and right panels are from  two different male speakers.]{
\begin{minipage}[c]{0.5\linewidth}
\centering
\includegraphics[width=4cm]{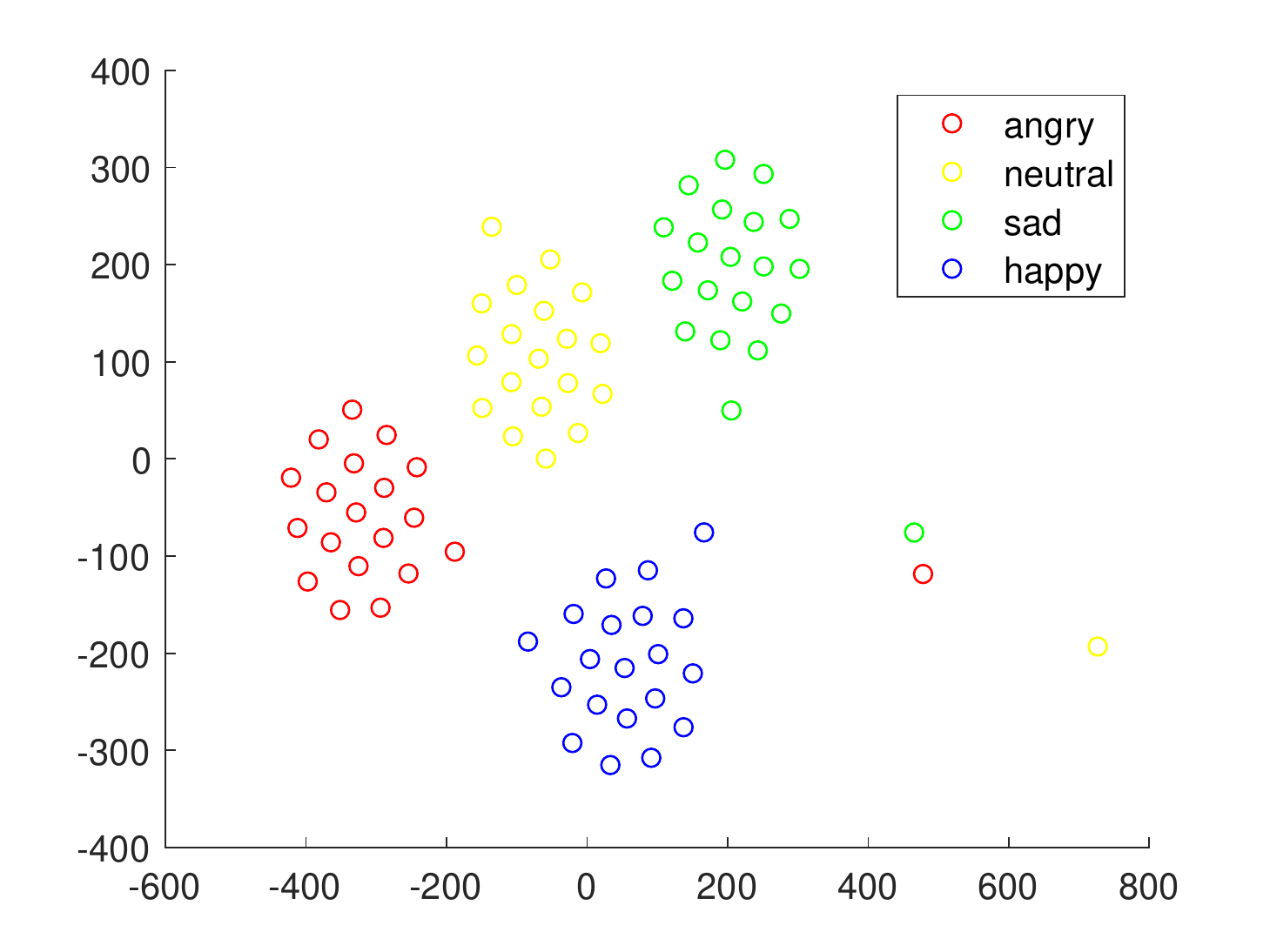}
\end{minipage}%
\begin{minipage}[c]{0.5\linewidth}
\centering
\includegraphics[width=4cm]{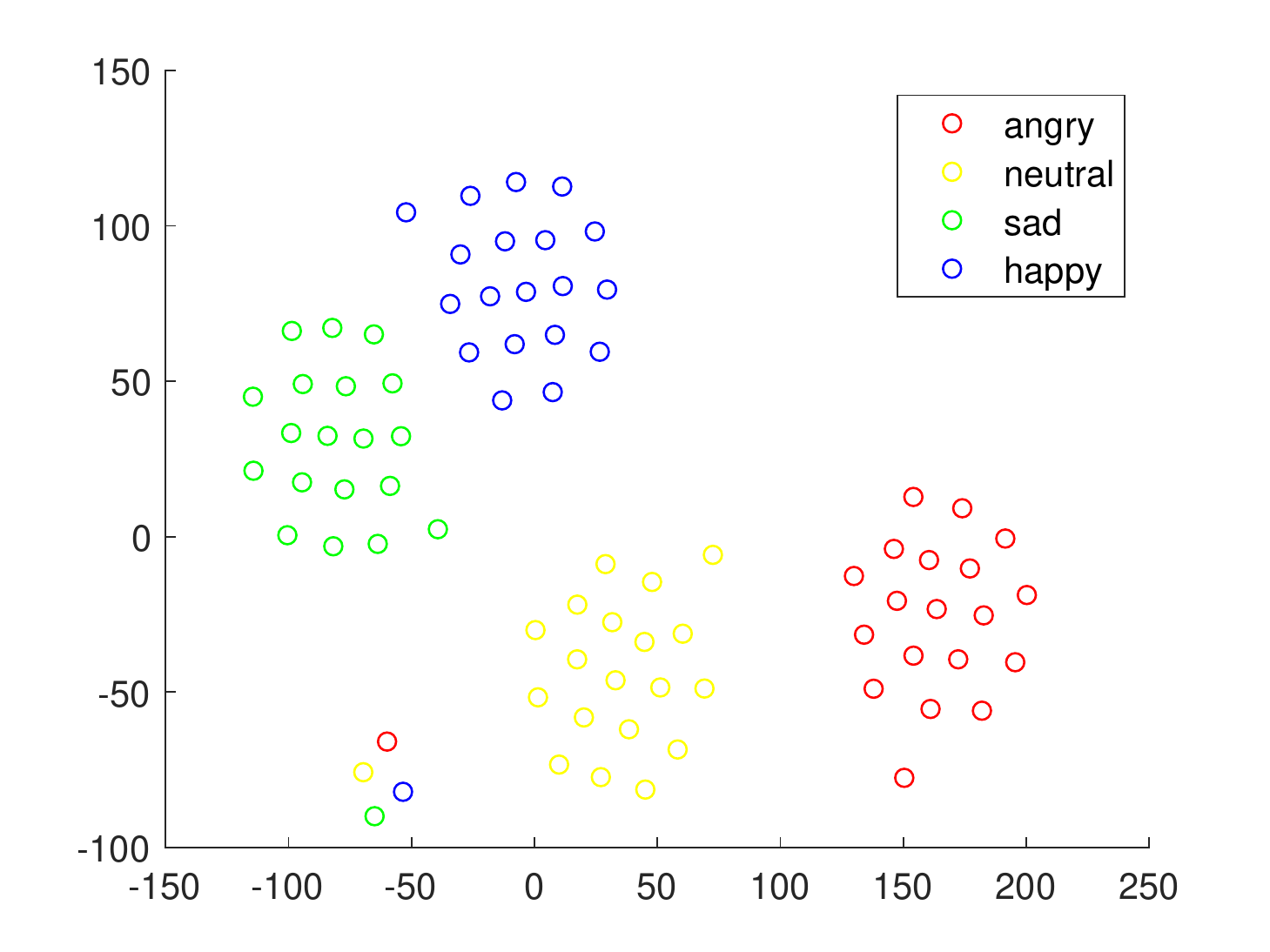}
\end{minipage}%
}%
\\
\vspace{-2mm}
\subfigure[Data in left and right panels are from  two different female speakers.]{
\begin{minipage}[c]{0.5\linewidth}
\centering
\includegraphics[width=4cm]{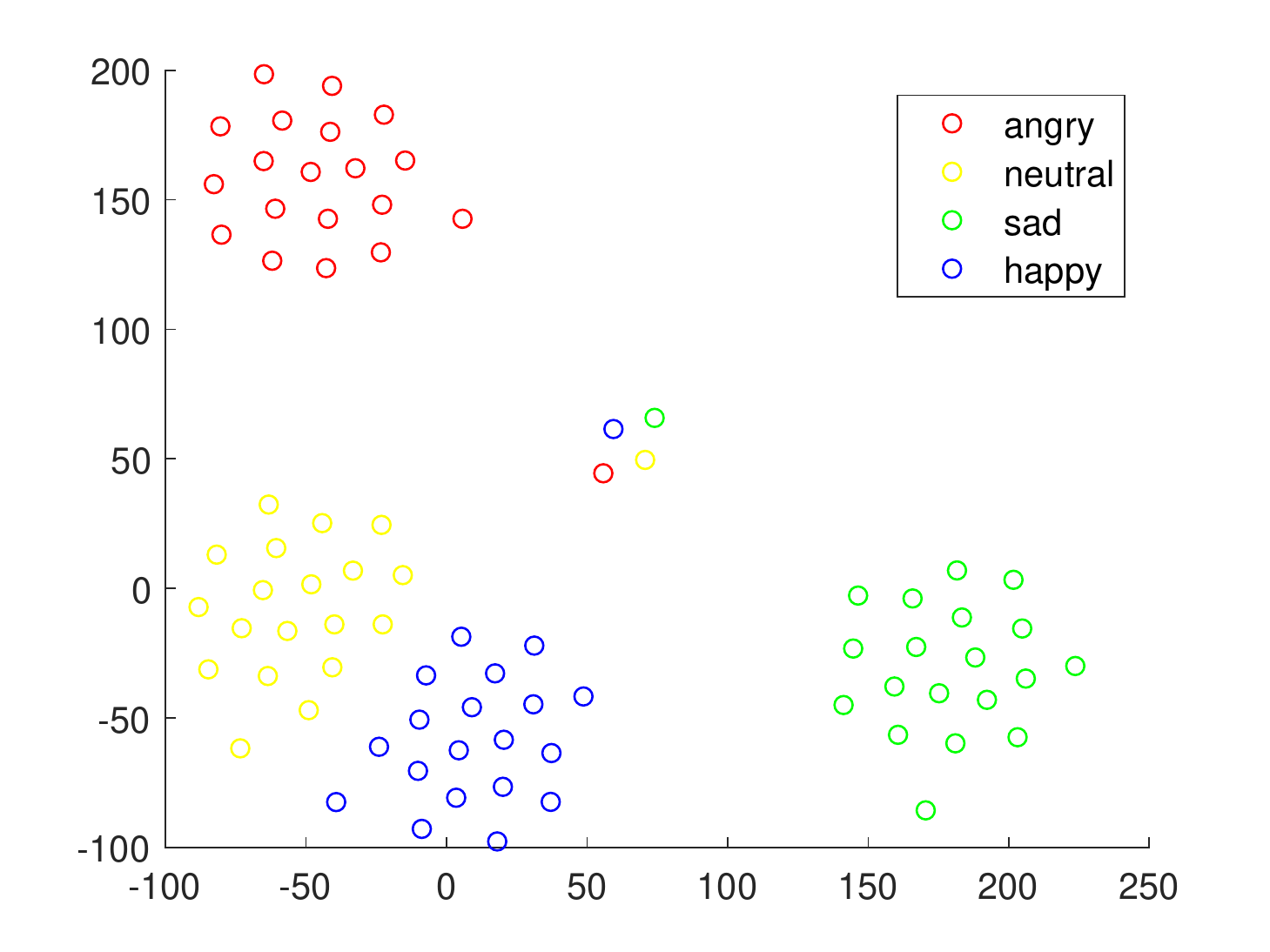}
\end{minipage}%

\begin{minipage}[c]{0.5\linewidth}
\centering
\includegraphics[width=4cm]{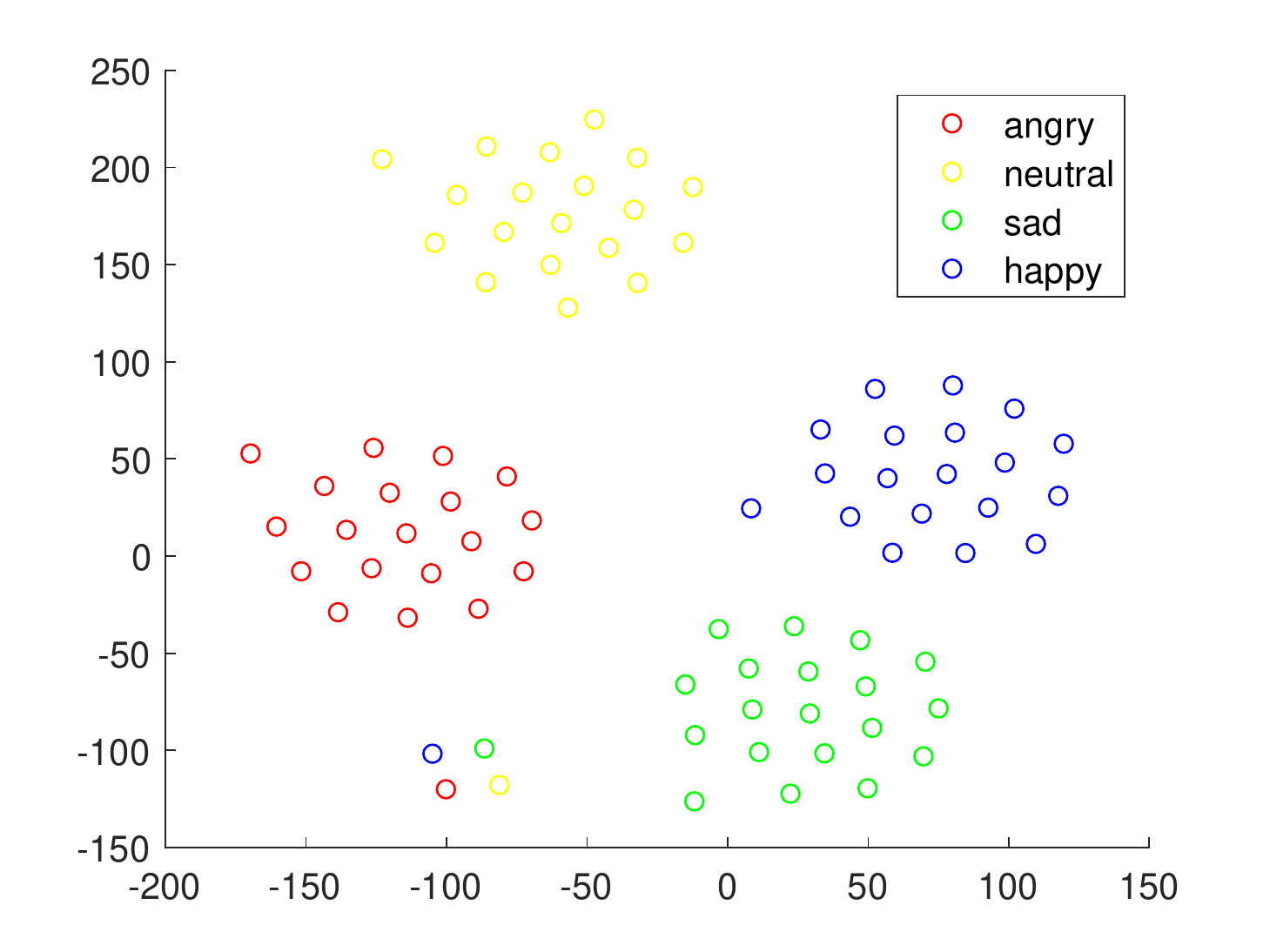}
\end{minipage}%
}%
\centering
\vspace{-2mm}
\caption{t-SNE plot of deep emotional features for 20 utterances with the same content but spoken by different speakers.}
\label{fig:visualization}
\vspace{-6mm}
\end{figure}

We are interested in the use of deep emotional features for voice conversion, to describe emotional prosody in a continuous space. The idea is to use deep emotional features of a reference speech to transfer its emotional style to an output target speech. To motivate the idea, we visualize the deep emotional features of 4 speakers (2 male and 2 female) using the t-SNE algorithm \cite{maaten2008visualizing} in a two-dimensional plane, as shown in Fig. \ref{fig:visualization}. It is observed that deep emotional features form clear emotion groups in terms of feature distributions. Fig. \ref{fig:visualization} suggests that we may use deep emotional features as a style embedding to encode an emotion class. Encouraged by this observation, we propose a one-to-many emotional style transfer framework through deep emotional features.


\section{One-to-many emotional style Transfer}
We propose a novel one-to-many emotional style transfer framework, that is based on VAW-GAN~\cite{hsu2017voice} with its decoder conditioning on deep emotional features.
The proposed one-to-many EST framework is referred as \textit{DeepEST} in short. We next discuss DeepEST in three stages: 1) emotion descriptor training, 2) encoder-decoder training with VAW-GAN, and 3) run-time conversion interface. In Stage \uppercase\expandafter{\romannumeral1}, we train an auxiliary SER network to serve as the emotion descriptor for input utterances. 
In Stage \uppercase\expandafter{\romannumeral2}, the proposed encoder-decoder training is implemented to learn the disentanglement and recomposition of the emotional elements. In Stage \uppercase\expandafter{\romannumeral3}, DeepEST takes input utterance and target deep emotional features to generate the utterance with target emotional style.

\subsection{Stage I: Emotion Descriptor Training}
Emotional prosody is complex with multiple acoustic attributes which makes it difficult to model. There have been studies to label emotion manually into discrete categories, such as one-hot emotion label \cite{busso2008iemocap, zhou2020converting}. As emotional prosody naturally spreads over a continuum that is hard to force-fit into a few categories, we propose to use deep emotional features that are learned from large animated and emotive speech data.

We propose to use a SER model as an emotion descriptor $D$, which extracts deep emotional features $\Phi$ from the input utterance $X$, or $\Phi = D(X)$. The SER architecture is as the same as that in \cite{chen20183}, which includes: 1) a three-dimensional (3-D) CNN layer; 2) a BLSTM layer; 3) an attention layer; 4) a fully connected (FC) layer. The 3-D CNN first projects the input Mel-spectrum with its delta and delta-deltas features into a fixed size latent representation, that preserves the effective emotional information while reducing the influence of emotional irrelevant factors. Then the following BLSTM and attention layer summarize the temporal information from the previous layer and produce discriminative utterance-level feature $\Phi$ for emotion prediction, as visualized in Fig. 1. 

\begin{figure*}
    \centering
    \includegraphics[width=14cm]{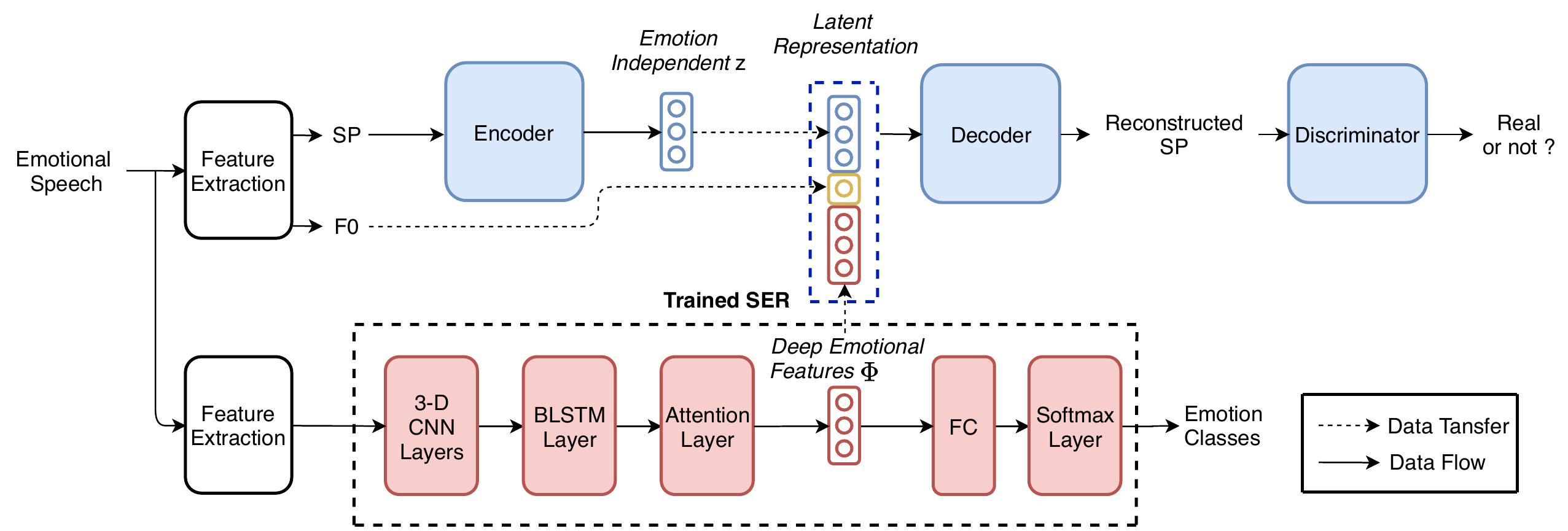}
   \vspace{-3mm}
    \caption{The training phase of the proposed {DeepEST} framework. Blue boxes represent the networks that involved in the training and red boxes represent the networks that are already trained. }
    \vspace{-3mm}
    \label{fig:training}
    \vspace{-3mm}
\end{figure*}

\begin{table*}[t]
\centering
\caption{MCD values of the baseline framework VAW-GAN-EVC and the proposed framework DeepEST in a comparative study. }
\scalebox{0.9}{
\begin{tabular}{c||c|c| c||c|c|c}
\hline
\multirow{2}{*}{MCD {[}dB{]}} & \multicolumn{3}{c||}{Male}          & \multicolumn{3}{c}{Female}         \\ \cline{2-7} 
                              & Zero Effort & VAW-GAN-EVC & \textbf{DeepEST}  & Zero Effort & VAW-GAN-EVC & \textbf{DeepEST}  \\ \hline 
neutral-to-happy                           & 6.769       & 4.738       & 4.569   & 7.088       & 4.284      & 4.260   \\ 
neutral-to-sad                           & 6.306       & 4.284       & 4.127   & 8.287       &  5.464       & 4.916   \\ 
\hline
neutral-to-angry                           & 6.649       & 4.482       & 4.564 \textit{(unseen)}  & 6.690      & 4.204       & 4.451  \textit{(unseen)}  \\ \hline
\end{tabular}}
\label{tab:mcd table}
\vspace{-5mm}
\end{table*}


\vspace{-3mm}
\subsection{Stage II: Encoder-Decoder Training with VAW-GAN}

Encoder-decoder structure has been used to effectively learn disentangled representations in previous studies \cite{zhou2020converting, hsu2017voice}. We propose an encoder-decoder training procedure as shown in Fig. \ref{fig:training}, where the encoder ($E$) learns to disentangle the emotional elements from the input features and generate a latent representation $z$. The resulting representation $z$ is assumed to contain phonetic and speaker information, but emotion-independent. Then the decoder/generator ($G$) learns to reconstruct the input features with the emotion-independent representation $z$ and other controllable emotion-related attributes.

In practice, we use WORLD vocoder \cite{morise2016world} to extract spectral features (\textit{SP}) and fundamental frequency ($F_0$) from the waveform. The encoder ($E_{\theta}$) with parameter set $\theta$ is exposed to the input spectral frames $x$ with different emotion types and learns an emotion-independent representation $z$: $z = E_{\theta}(x)$. Since the latent representation $z$ extracted from the source spectrum still contains the source $F_0$ information, and the conversion performance can suffer from this flaw \cite{zhou2020converting}. Therefore, the decoder/generator ($G_{\psi}$) with parameter set $\psi$ takes emotion-independent representation $z$ and the emotion-related features: the deep emotional features $\Phi$ that reflect the global emotional variance of the input utterance $X$ from Stage I and the corresponding $F_0$ that contains source pitch information to recompose the emotional elements of the spectrum. The reconstructed feature $\overline{x}$ can be formulated as:
\begin{equation}
    \overline{x} = G_{\psi}(z,\Phi,F_0) = G_{\psi}(E_{\theta}(x),D(X), F_0)
\end{equation}

We then train a generative model for spectrum through an adversarial training: The discriminator ($Y_{\mu}$) with parameter set $\mu$ tries to maximize the loss between the real features $x$ and reconstructed features $\overline{x}$, while the generator ($G_{\psi}$) tries to minimize it. The parameter sets $\theta$, $\psi$ and $\mu$ are optimized through this min-max game, which allows us to generate high-quality speech samples.
\vspace{-3mm}
\subsection{Stage III: Run-time Conversion}

During the run-time conversion, we have a source utterance that is expressed in neutral emotion, we would like to convert it to a target emotion following the reference emotion style from the reference utterances. 
Suppose that we have a set of reference utterances $X_t$ belonging to an emotion category. 
We first use the pre-trained SER to generate the deep emotional features $\Phi_t = mean(D(X_t))$ for all the reference utterances, i.e. all the utterances of our dataset with the same reference emotion. We then concatenate $\Phi_t$ with the converted $F_0$ ($\hat{F_0}$) and emotion-independent $z$ from the source utterance to compose a latent representation for the target utterance \textit{SP}. The converted \textit{SP} can be formulated as:
\begin{equation}
    \hat{x} = G_{\psi}(z, \Phi_t, \hat{F_0}) = G_{\psi}(E_{\theta}(x),mean(D(X_t)), \hat{F_0}) 
\end{equation}
Finally, the converted speech with target emotion is synthesised by WORLD vocoder with converted spectral features and converted $F_0$.


\section{Experiments}

\subsection{Emotional Speech Dataset (ESD) }
With this paper, we introduce and publicly release a new multi-lingual and multi-speaker emotional speech dataset that can be used for various speech synthesis and voice conversion tasks\footnote{{\textbf{Emotional Speech Dataset (ESD):} \url{https://github.com/HLTSingapore/Emotional-Speech-Data}}}.
The dataset consists of 350 parallel utterances with an average duration of 2.9 seconds spoken by 10 native English and 10 native Mandarin speakers. For each language, the dataset consists of 5 male and 5 female speakers in five emotions summarized as follows: 1) happy, 2) sad, 3) neutral, 4) angry, and 5) surprise. Speech data are sampled at 16 kHz and saved in 16 bits. 

To our best knowledge, this is the first parallel voice conversion dataset that provides emotion labels in a multi-lingual and multi-speaker setup. As a future work, we will report an in-depth investigation of this dataset for cross-lingual and mono-lingual emotional voice conversion applications. 
\vspace{-4mm}
\subsection{Experimental Setup}

We conduct objective and subjective evaluation to assess the performance of our proposed DeepEST model for seen and unseen emotional style transfer. We use four English speakers (2 male and 2 female) from ESD  dataset. For each speaker, we conduct \textit{seen emotion} conversion from neutral to happy (N2H: neutral-to-happy) and neutral to sad (N2S: neutral-to-sad). We choose angry as the \textit{unseen emotion}, and conduct experiments from neutral to angry (N2A: neutral-to-angry) to assess the performance of our proposed model for unseen emotion style transfer. 

We split the 350 utterances in ESD dataset for a speaker into training set (300 utterances), reference set (30 utterances), and test set (20 utterances). During training, we propose to have \textbf{one universal model} that takes 300 utterances from neutral, happy and sad emotion states respectively. To obtain the deep emotional features, we train the SER \cite{gao2020interactive, liu2020expressive} on a subset of IEMOCAP \cite{busso2008iemocap} with four emotion types (happy, angry, sad and neutral).  At run-time, we evaluate our framework with 20 utterances both from seen (happy, sad) and unseen (angry) emotion states. 
We obtain the reference emotion style by using the 30 utterances from the reference set, as formulated in Eq. (2). 
For each emotion conversion, we generate the deep emotional features from SER module by calculating the mean of the features that are in the same emotion category as the emotion reference.

As the baseline framework, we adapted a state-of-the-art {VAW-GAN-EVC} \cite{zhou2020converting} that can perform conversion from one emotional state to another. We note that for each emotion conversion pair, we need to train a new VAW-GAN-EVC model as it is not capable of performing one-to-many conversion.  In contrast, DeepEST provides more flexible manipulation and generation of the output emotion, as it can perform an emotional conversion from one emotional state to many and generate unseen emotions, which will be investigated in the following sections.

\begin{table}[t]
\centering
\caption{MOS results with 95 \% confidence interval to assess the speech quality.}
\scalebox{0.9}{

\begin{tabular}{c||c|c|c}
\hline
MOS & N2H & N2S & N2A \\ \hline
Reference & 4.95 $\pm$ 0.11    &  4.88 $\pm$ 0.22     & 4.87 $\pm$ 0.22         \\ 
VAW-GAN-EVC & 3.23 $\pm$ 0.71   & 2.80 $\pm$ 0.55     & 3.11 $\pm$  0.57         \\ 
\textbf{DeepEST} & 3.24 $\pm$ 0.72   & 2.94 $\pm$ 0.57      & 3.15 $\pm$ 0.63           \\ \hline
\end{tabular}}
\label{tab:mos}
\vspace{-5mm}
\end{table}

In DeepEST, 513-dimensional \textit{SP}, $F_0$ and \textit{APs} are extracted every 5 ms with the FFT length of 1024 using the WORLD vocoder. The frame length is 25 ms with a frame shift of 5 ms. We normalize every input frame of \textit{SP} to unit sum and then re-scale it to logarithm. The encoder is a 5-layer 1D CNN with the kernel size of 7 and a stride of 3 followed by a FC layer. Its output channel is $\{$16,32,64,128,256$\}$. The latent representation $z$ is 128-dimensional and assumed to have a Gaussian distribution. The deep emotional feature is 256-dimensional, and concatenated with the 128-dimensional latent representation $z$ and 1-dimensional $F_0$ contour to merge as the input to the decoder. The decoder is 4-layer 1D CNN with kernel size of $\{$9,7,7,1025$\}$ and strides of $\{$3,3,3,1$\}$. Its output channel is $\{$32,16,8,1$\}$. The discriminator is a 3-layer 1D CNN with kernel size of $\{$7,7,115$\}$ and strides of $\{$3,3,3$\}$ followed by a FC layer. The networks are trained by using RMSProp with a learning rate of 1e-5. The batch size is set as 256 for 45 epochs.



\vspace{-3mm}
\subsection{Objective Evaluation}
We conduct objective evaluation to assess the performance of our proposed model. We calculate Mel-cepstral distortion (MCD) \cite{sisman2019group,sisman2020overview} to measure the spectral distortion between the converted and reference Mel-spectrum for two male and two female speakers for three emotion combinations.


As shown in Table \ref{tab:mcd table}, the proposed DeepEST outperforms the baseline framework VAW-GAN-EVC for all the seen emotions (N2H and N2S). We note that for the unseen emotion (\textit{angry}), DeepEST still achieves comparable results to that of VAW-GAN-EVC baseline, which is trained with \textit{angry} speech samples. These encouraging observations validate the effectiveness of our proposed DeepEST for seen and unseen emotion transfer. Last but not least, we require three VAW-GAN-EVC models to be trained for all conversion pairs, whereas the proposed DeepEST can perform all the emotion mapping pairs within one model,  which we believe is remarkable.
\vspace{-3mm}
\subsection{Subjective Evaluation}

We further conduct three listening experiments to assess the proposed DeepEST in terms of speech quality and emotion similarity. 15 subjects participated in all the experiments and each listened to 108 converted utterances in total.

\begin{figure}[t]
    \centering
    \includegraphics[width=8cm]{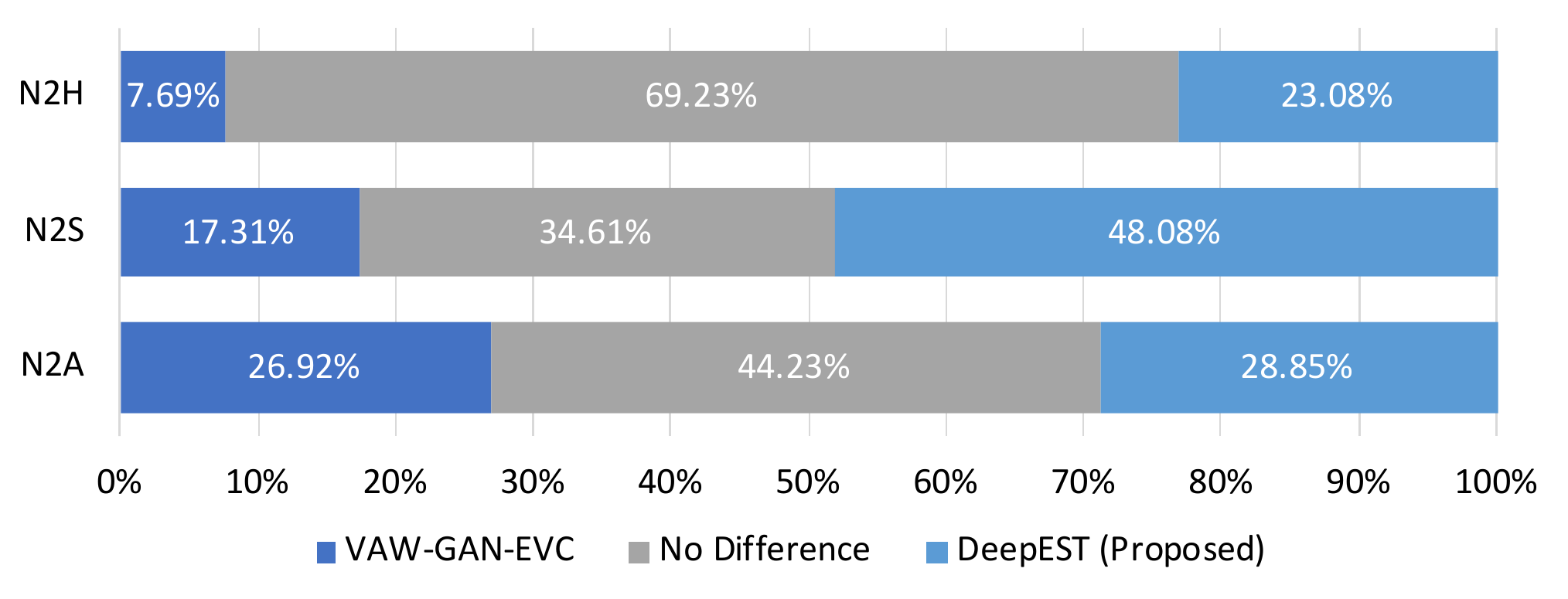}
    \vspace{-3mm}
    \caption{AB preference test results for the speech quality. }
    \label{fig:xab_1}
    \vspace{-3mm}
\end{figure}

We first report the mean opinion score (MOS) of the reference speech samples, baseline VAW-GAN-EVC and the proposed  DeepEST. As shown in Table \ref{tab:mos}, DeepEST achieves better results than the baseline for both seen and unseen emotion combinations, which we believe is remarkable. Secondly, we conduct AB preference test to further evaluate the speech quality, where the subjects are asked to choose the speech samples with higher speech quality.  As shown in Fig. \ref{fig:xab_1}, we observe that proposed framework DeepEST consistently outperforms the baseline framework VAW-GAN-EVC for all the emotion combinations, which is also consistent with the MOS results. These observations validates the effectiveness of our proposed model in terms of voice quality. 

\begin{figure}[t]
    \centering
    \includegraphics[width=8cm]{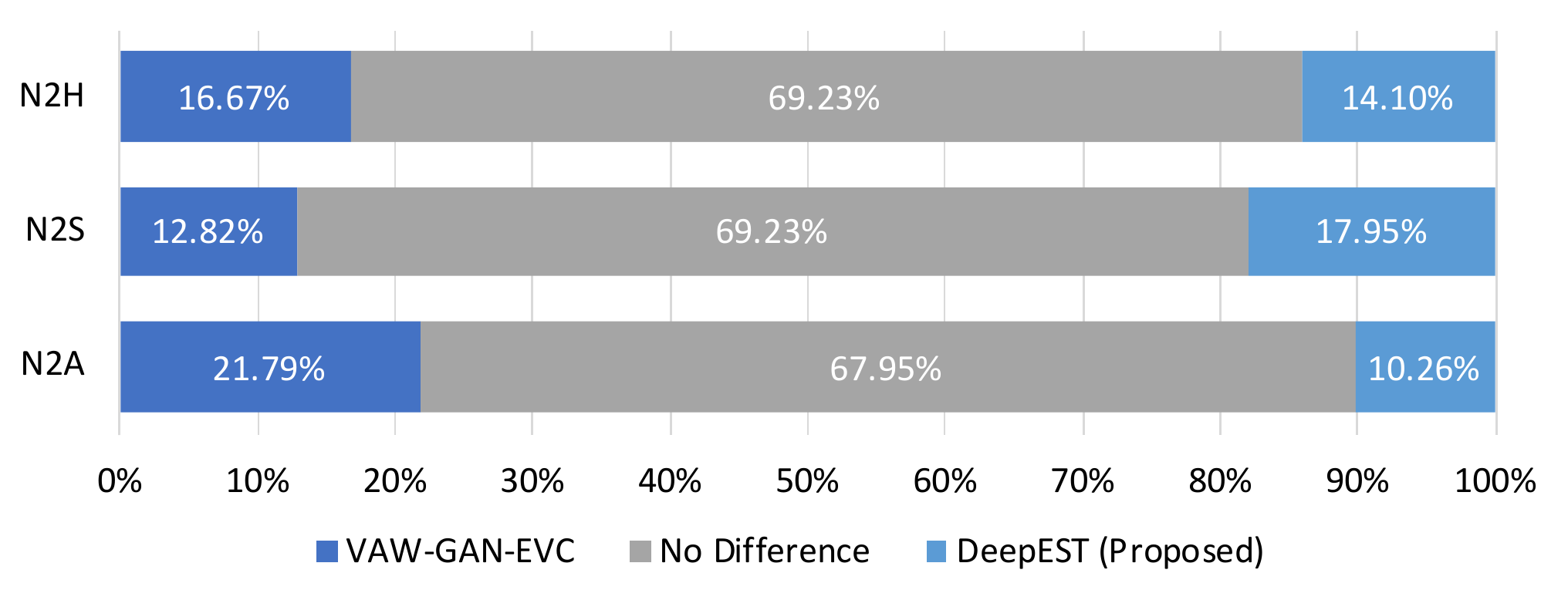}
    \vspace{-5mm}
    \caption{XAB preference test results for the emotion similarity.}
    \vspace{-4mm}
    \label{fig:xab_2}
\end{figure}

We further conduct XAB emotion similarity test to assess the emotion conversion performance, where subjects are asked to choose the speech samples which sound closer to the reference in terms of the emotional expression. Consistent with previous experiments, the baseline is one-to-one conversion and all emotions are seen emotions. As shown in Fig. \ref{fig:xab_2}, DeepEST outperforms the baseline for neutral-to-sad conversion. We observe that baseline achieves better performance for neutral-to-happy conversion due to the poor performance of SER on happy (29.95\%) compared with 84.32\% on sad, and 70.47\% on angry. For unseen emotion (\textit{angry}), DeepEST still achieves comparable results with the baseline in terms of emotion similarity, which is very encouraging. This result validates the effectiveness of DeepEST for unseen emotion transfer. As a future work, we will improve the SER performance for all emotional states, and report an in-depth investigation. 
\vspace{-3mm}


\section{Conclusion}
In this paper, we propose a one-to-many emotional style transfer framework based on VAW-GAN without the need for parallel data. We propose to leverage deep emotional features from SER to describe emotional prosody in a continuous space. By conditioning the decoder with controllable attributes such as deep emotional features and F0 values, we achieve competitive results for both seen and unseen emotions over the baseline framework, which validates the effectiveness of our proposed framework. Moreover, we also introduce a new emotional speech dataset, ESD, that can be used in speech synthesis and voice conversion.
\section{Acknowledgment}
The research is supported by the National Research Foundation, Singapore under its AI Singapore Programme (Award No: AISG-GC-2019-002) and (Award No: AISG-100E-2018-006), and its National Robotics Programme (Grant No. 192 25 00054), and by RIE2020 Advanced Manufacturing and Engineering Programmatic Grants A1687b0033, and A18A2b0046. The research by Berrak Sisman and Rui Liu is funded by SUTD Start-up Grant Artificial Intelligence for Human Voice Conversion (SRG ISTD 2020 158) and SUTD AI Grant, titled 'The Understanding and Synthesis of Expressive Speech by AI'. 

\newpage
\bibliographystyle{IEEEbib}
{\footnotesize
\bibliography{mybib}
}
\end{document}